# Dynamics of Interfacial Diffusion Control in Amphiphilic Lipid-Coated Micro-Particles for Stochastic Release Systems

[1],*Joonggwon Kim, [2],**Dogyun Byeon
*unexa301@gmail.com
**dogun04@changwon.ac.kr

(Date : December 31, 2025)

[1]UNEXA Korea Inc., 3903, A, 97 Centum jungang-ro, Haeundae-gu, Busan, 48058, Republic of Korea
[2]Institute of Advanced Nano Science and Technology, Changwon National University, Changwon, 51140, Republic of Korea

**Abstract**

The release of hydrophilic solutes from micron-scale particulate formulations can be understood as an interfacial transport problem in which diffusion across a heterogeneous amphiphilic coating competes with dissolution and convective removal in the surrounding medium. Here we reinterpret a glycerin fatty acid ester (GFAE)–coated thiamine (vitamin B1) micro-particle formulation as a condensed-matter system: a soft-matter core–shell geometry whose effective permeability is set by the nanoscale organization of amphiphilic lipids at the interface. Using in vivo time-course serum measurements in mice as a proxy for a stochastic sink, we compare the coated formulation (UTEV) with a composition-matched uncoated comparator (UMFG). Early-time systemic appearance is similar, whereas late-time levels are enhanced for the coated particles, implying a reduced effective interfacial diffusivity and a broadened release-time distribution. We discuss the results in terms of diffusion-barrier physics, heterogeneous interfacial energetics, and coarse-grained transport models that map microstructural coating parameters to macroscopic persistence (AUC).

## 1. Introduction

In the functional food and nutraceutical industries, high absorption is often promoted as a key indicator of product performance. However, pharmacokinetic considerations suggest that an excessively rapid absorption profile is not universally beneficial for water-soluble micronutrients. When systemic levels rise abruptly after oral intake, efficient renal clearance can shorten the duration of meaningful exposure, resulting in a steep peak-and-decline profile and potentially reducing practical supplementation efficiency when persistence is limited [1]. Accordingly, formulation strategies are increasingly shifting from maximizing early uptake to optimizing physiological maintenance—i.e., sustaining more stable systemic availability over time while avoiding unnecessary concentration spikes. From a transport perspective, this situation can be viewed as a sink-dominated boundary condition, where rapid clearance continually removes solute from circulation and effectively emphasizes the importance of time-dependent input rather than peak concentration alone.

Intestinal absorption of water-soluble vitamins involves regulated and specific mechanisms that can influence systemic exposure profiles [2]. Thiamine (vitamin B1) is a compelling target for persistence-oriented delivery. Thiamine is essential for carbohydrate metabolism and energy production through its biologically active form, thiamine pyrophosphate (TDP), which serves as a cofactor for multiple enzymes. Recent mouse studies also suggest that thiamine supplementation can influence metabolic phenotypes, underscoring the physiological relevance of sustained thiamine availability [2]. Classical pharmacokinetic studies in humans indicate that thiamine can appear rapidly in systemic circulation after oral administration but may also be cleared quickly, resulting in short persistence despite high intake levels. From a formulation standpoint, therefore, the central challenge is not simply to increase absorption, but to prolong effective availability under physiological conditions. At a coarse-grained level, systemic exposure can be considered as the combined consequence of formulation-controlled release and sink-side removal kinetics; thus, broadening release in time is expected to reduce unnecessary early overshoot while extending meaningful mid-to-late availability.

In addition to rapid clearance, thiamine presents stability challenges during processing, storage, and gastrointestinal transit. Water-soluble vitamins can be susceptible to oxidation, heat, and pH-dependent degradation, which may reduce delivered dose and contribute to variability in bio-accessibility [3]. To address such limitations, lipid-based delivery systems—including liposomes, nanoemulsions, and lipid nanocarriers—have gained attention for protecting sensitive actives and modulating release behavior [4–7]. Importantly, although thiamine is hydrophilic, several reports suggest that edible lipid-based encapsulation can improve thiamine handling by engineering microenvironments or interfaces that reduce exposure to destabilizing conditions and enable controlled release [3,8–10]. Similar nanoencapsulation strategies have been applied to other vitamins (e.g., vitamin A) to improve stability and oral delivery [11]. These approaches provide a practical route for converting a rapid rise/rapid fall profile into a more sustained exposure pattern. In physical terms, controlled release from coated particles represents a diffusion- and partitioning-governed transport process, where a solute initially confined in a solid core traverses a heterogeneous amphiphilic barrier whose microstructure is shaped by interfacial energetics and phase behavior.

Mechanistically, intestinal thiamine uptake is mediated by specialized transport processes, including the human thiamine transporter hTHTR-2, which contributes to regulated absorption in the small intestine [4]. A sustained-release strategy that gradually presents thiamine for absorption—rather than delivering it as an immediate bolus—may help reduce excessive peak concentrations while extending the duration of measurable systemic levels. In this context, sustained-release formulations aim to improve time-dependent maintenance and overall exposure rather than emphasizing early peak responses. Within this framework, a lipid coating can be viewed as an interfacial permeability regulator that moderates particle-to-fluid flux while still allowing transporter-mediated uptake to proceed over an extended time window.

Here, we developed a lipid-coated thiamine formulation (UTEV) using a glycerin fatty acid ester (GFAE) coating strategy. GFAE is an amphiphilic coating material capable of forming a robust, lipophilic protective layer on the surface of thiamine particles. UTEV was produced via a fluidized-bed coating process to achieve uniform deposition. This surface-engineering approach is designed to protect thiamine during gastrointestinal transit and to modulate dissolution and release kinetics, thereby supporting sustained systemic maintenance. To our knowledge, the in vivo time-course persistence of a GFAE-coated thiamine powder has been less explored in direct comparison with a composition-matched formulation. We therefore interpret serum time courses using a minimal transport viewpoint that conceptually links coating-controlled release to effective diffusional permeability and, in turn, to systemic maintenance under sink-dominated clearance.

To evaluate in vivo performance, we conducted a comparative time-course study in male C57BL/6 mice. Mice received a single oral dose (1 mg/kg) of UTEV or a composition-matched comparator (UNEXA matched formula group; UMFG), and serum vitamin B1 concentrations were quantified at multiple time points spanning early, mid, and late phases after dosing using an ELISA-based assay. We additionally assessed baseline-normalized profiles and overall exposure via area-under-the-curve (AUC) analysis. We hypothesized that lipid coating would primarily enhance mid-to-late maintenance—yielding comparable early post-dose levels but higher later concentrations and increased integrated exposure relative to UMFG—supporting a shift from absorption-centric to persistence-centric thiamine delivery.

## 2. Transport Model and Physical Picture

We model sustained release from coated micro-particles as an interfacial transport problem in a heterogeneous amphiphilic barrier. Each particle is treated as a core–shell geometry in which diffusion through the shell competes with dissolution and removal by an external sink. Shell microstructure (e.g., tortuosity, transient free volume, and micro-phase separation) controls the effective diffusivity $D_{eff}$, producing a distribution of release times across an ensemble. The measured serum concentration further reflects convolution of the release kernel with systemic clearance.

### 2.1. Core–shell permeability and characteristic time scale

We represent a micro-particle as a spherical core (radius $R$) containing a soluble species at initial concentration $c_0$, surrounded by an amphiphilic shell of thickness $\delta$. In the steady-flux limit, transport across the shell is governed by the permeability

$$P = \frac{D_{eff}}{\delta}.$$

The outward molar flux can be written as

$$J \simeq 4\pi R^2 P (c_{core} - c_{ext}),$$

where $c_{ext}$ is the concentration in the surrounding medium. For sink-driven release ($c_{ext} \approx 0$), the characteristic release time scale scales as

$$\tau \sim \frac{\delta R}{D_{eff}}.$$

Variability in $\delta$ and $D_{eff}$ across particles therefore yields a stochastic distribution of $\tau$, broadening the ensemble release kinetics.

### 2.2. Tortuosity-controlled effective diffusivity

In amphiphilic lipid–glycerin shells, solute transport occurs through a percolating hydrophilic network whose geometry is set by micro-phase separation and interfacial packing. A standard coarse-grained mapping relates the effective diffusivity to the free-solution value D0 through

$$D_{eff} = D_0 \left(\frac{\varepsilon}{\tau^2}\right) K,$$

where ε is the volume fraction of transport-accessible (hydrophilic) pathways, τ is pathway tortuosity, and K is a partition factor capturing energetic preference at the amphiphilic interface. For labyrinthine channels, $\tau \approx \varepsilon^{-\alpha}$ with $\alpha \sim 0.5 - 1$, giving $D_{eff} \approx D_0\, \varepsilon^{1+2\alpha} K$. Increased lipid packing reduces ε and increases τ, thereby lowering D_eff and permeability P. Particle-to-particle variability in $\delta$, $\varepsilon$ and $\tau$ produces a distribution of permeabilities, naturally generating broad release-time statistics even when the mean shell fraction is fixed.

### 2.3. Stochastic release and CTRW in disordered shells

Shell microstructure can be treated as a disordered landscape of hydrophilic domains separated by lipid-rich barriers. Transport can thus be viewed as intermittent hopping between domains. If waiting times between successive hops are broadly distributed, release becomes anomalous and is well described by a continuous-time random walk (CTRW). Let $\psi(t)$ be the waiting-time distribution for a hop event. For a heavy-tailed form $\psi(t) \sim t^{-1-\beta}$ with $0 < \beta < 1$, the mean waiting time diverges and first-passage (release) times develop a long-time power-law tail. At the ensemble level, the release kernel $r_{rel}(t)$ exhibits a stretched early-time regime followed by an algebraic tail. Operationally, this corresponds to a non-negligible fraction of particles acting as long-lived traps, producing an extended late-time contribution to serum concentration and increasing AUC without amplifying the early peak.

### 2.4. Coupling to systemic clearance

In vivo, the measured serum concentration $C(t)$ is shaped by the release kernel and systemic clearance with rate $k_{cl}$. A minimal linear-response representation is

$$C(t) = \int_0^t r_{rel}(t') \exp[-k_{cl}(t - t')] \, dt'.$$

In this framework, enhanced late-time $C(t)$ without increased early $C(t)$ is consistent with reduced $D_{eff}$ (or increased $\delta$) that shifts $r_{rel}(t)$ toward longer times or generates heavier-tailed release-time statistics, while preserving total released mass.

## 3. Materials and Methods

### 3.1. Particle formulation and interfacial coating

Two test articles were used: (i) UMFG, a composition-matched comparator, and (ii) UTEV, a GFAE-coated thiamine micro-particle formulation. GFAE is an amphiphilic ester comprising a hydrophilic glycerin moiety and lipophilic fatty-acid chains. A coating solution was prepared by esterification (75.5% stearic acid, 24.5% glycerin, w/w) and dissolution in ethanol. Coating was applied onto thiamine-HCl cores by fluidized-bed processing to produce a nominal 35% (w/w) shell fraction (core:shell = 65:35) (**Fig. 1**)[12].

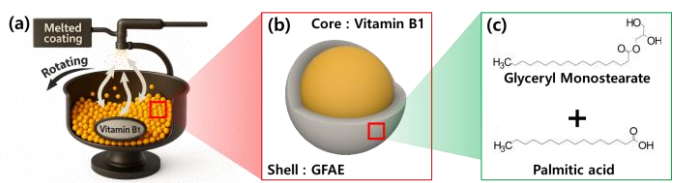

**Fig. 1**. Schematic of the core–shell micro-particle architecture and coating process.

A hydrophilic thiamine core is coated by an amphiphilic GFAE shell of thickness $\delta$. Release is governed by partitioning at the core–shell interface and diffusion through a heterogeneous shell microstructure characterized by porosity $\varepsilon$ and tortuosity $\tau$, producing $D_{eff}$ and permeability $P = D_{eff}/\delta$.

GFAE, an amphiphilic food-additive compound bearing a hydrophilic glycerin moiety and lipophilic fatty acid chains, was selected to form a uniform lipid-friendly protective layer on the surface of water-soluble vitamin particles. For preparation of the coating material, GFAE was produced by an esterification reaction based on a formulation consisting of 75.5% fatty acid (=stearic acid) and 24.5% glycerin (w/w), and dissolved in an ethanol to generate a coating solution suitable for particle coating. UTEV was produced by applying the GFAE coating solution onto the water-soluble vitamin B1 raw material (=thiamine HCl) using a fluidized-bed coating technique to achieve uniform deposition of the lipid layer[13]. The formulation composition was 65% vitamin core and 35% GFAE coating (w/w). This encapsulation strategy was designed to protect vitamin B1 from gastrointestinal degradation and to modulate dissolution and absorption kinetics, thereby supporting a sustained systemic concentration profile. The key specifications are summarized in **Table 1**.

**Table 1**. Composition and key specifications of the comparator (UMFG) and the lipid-coated formulation (UTEV).

| UNEXA matched formula group ; UMFG | | UNEXA technology-enhanced vitamin ; UTEV |
|---|---|---|
| Thiamine HCl | Core material | Thiamine HCl |
| 100 | Vitamin B1 content (%) | 65 |
| None | Coating material | GFAE |
| 0 | Coating ratio (%) | 35 |
| 100-500 | Particle size (μm) | 100-500 |

### 3.2. Animals, husbandry, and ethics

Male C57BL/6 mice were used for the in vivo time-course evaluation of serum vitamin B1 levels. Animals were purchased at 7 weeks of age (approximately 19–24 g) and acclimatized for 5 weeks prior to study initiation. At the time of dosing, mice were 12 weeks old and weighed approximately 26–33 g. Animals were maintained under specific pathogen-free (SPF) conditions with controlled environmental parameters, including a temperature of 23 ± 2°C, relative humidity of 55 ± 15%, noise levels below 60 phon, and a 12 h light/dark cycle (lights on from 08:00 to 20:00) with illumination maintained at 150–300 lux. Air exchange was regulated at 10–12 changes per hour. Throughout acclimatization and experimentation, mice were provided a standard pellet diet (Purina Lab Rodent Chow #38057, Purina Co., Seoul, Korea) and filtered drinking water ad libitum (water replaced daily). All animal procedures were conducted in accordance with Institutional Animal Care and Use Committee (IACUC) guidelines, and the study protocol was approved under identifier IV-RA-27-2504-09 (Invivo Co., Ltd.).

### 3.3. Study design, randomization, and oral administration

Following acclimatization, mice were weighed and assigned to experimental groups using a randomized block design to ensure comparable mean body weights between groups. Individual animals were identified by ear punching. Two experimental groups were established (n = 8 per group): UMFG and UTEV. Test materials were freshly prepared in distilled water on the day of administration. Both formulations were administered orally at a dose of 1 mg/kg, normalized to thiamine HCl content (1 mg/kg as vitamin B1 equivalent). Accordingly, the administered mass of UTEV was adjusted to account for its 65% vitamin content. Animals were fasted for ≥12 h prior to dosing to minimize variability arising from feeding status. Blood sampling was performed at predefined time points (0, 3, 6, and 24 h) to probe time-dependent serum vitamin B1 profiles and to capture formulation-dependent release–clearance dynamics.

### 3.4. Blood collection, serum preparation, and storage

Blood samples were collected from the jugular vein at baseline (0 h, pre-dose) and at 3, 6, and 24 h after oral administration of UMFG or UTEV. Collected blood was transferred into microcentrifuge tubes and allowed to stand for 30 min to facilitate clot formation. Samples were then centrifuged at 3,000 rpm for 15 min at 4°C to isolate serum. The resulting serum was stored at −80°C until quantitative analysis.

### 3.5. ELISA quantification as a time-resolved sink assay

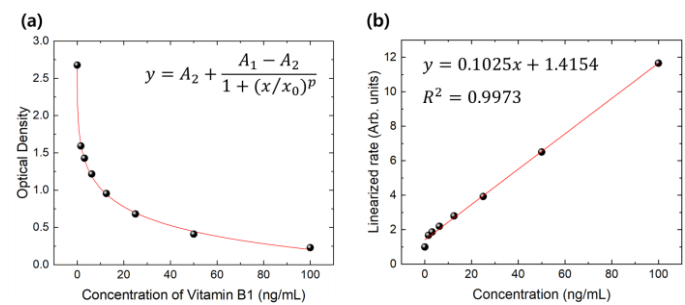

**Fig. 2**. ELISA calibration curve used to convert optical density to serum vitamin B1 concentration. (a) OD versus concentration fitted by a 4-parameter logistic model. (b) Linearized response for quality control and working-range assessment.

Serum vitamin B1 concentrations were quantified using a commercially available Mouse Vitamin B1 (V) ELISA kit (Catalog No. MOEB2555; Assay Genie, Dublin, Ireland) according to the manufacturer's protocol. After completion of the ELISA procedure, optical density (OD) values were converted to vitamin B1 concentrations via interpolation from a plate-specific calibration curve generated using the kit standards (Fig. 2). The OD–concentration relationship was fitted using a four-parameter logistic (4PL) model (Fig. 2(a)). To evaluate linearity across the working range, the OD data were additionally transformed into a linearized response and analyzed by linear regression (Fig. 2(b); $y = 0.1025x + 1.4154$, $R^2 = 0.9973$). Concentrations are reported in ng/mL at each sampling time point, and group data are summarized as mean ± S.E.. Although vitamin B1 pharmacokinetics can be assessed more directly by measuring thiamine diphosphate (TDP) in erythrocytes or whole blood via HPLC, the present study employed serum ELISA as a practical, time-resolved readout for comparing formulations. From a coarse-grained transport perspective, the serum time course can be viewed as an effective sink response: it reflects the convolution of the particle-scale release kernel with absorption and systemic clearance processes. As such, differences in the late-time tail of the serum profile provide a sensitive indicator of permeability-controlled release and broadened release-time statistics in heterogeneous coated particles [10]. In particular, an extended late-time tail is consistent with either reduced effective diffusivity (e.g., tortuosity-controlled transport through a heterogeneous shell) or stochastic trapping that generates heavy-tailed release-time distributions, as expected in CTRW-like dynamics.

### 3.6. Exposure assessment and statistical analysis

To compare systemic exposure between formulations over the observation window, systemic exposure was quantified by the area under the concentration–time curve, AUC(0–24 h), computed via the trapezoidal rule using measured concentrations at 0, 3, 6, and 24 h. AUC is a standard metric for comparing systemic exposure and bioavailability [14,15]. Statistical analyses were performed using SPSS version 23.0 (SPSS Inc., Chicago, IL, USA). Differences between UMFG and UTEV at corresponding time points were evaluated using two-sample t-tests, and p values < 0.05 were considered statistically significant. For relative (baseline-normalized) analyses, one-way analysis of variance (ANOVA) was conducted; when significant group effects were observed (p < 0.05), Duncan's multiple range test was applied for post hoc comparisons. Because early-time concentrations were comparable between formulations, differences in AUC(0–24 h) were expected to be driven primarily by the late-time tail of the concentration profile, consistent with permeability-controlled release and broadened release-time statistics in heterogeneous coated particles.

# 4. Results & Discussion

## 4.1. Time-dependent serum profiles reveal permeability-controlled persistence

Serum vitamin B1 concentration–time profiles were measured at baseline (0 h) and at 3, 6, and 24 h following a single oral administration (1 mg/kg, thiamine-HCl equivalent) of either the uncoated comparator UMFG or the amphiphilic lipid-coated formulation UTEV (n = 8 per group; Fig. 3). Baseline serum concentrations were comparable between groups (UMFG: 96.09 ± 2.52 ng/mL; UTEV: 93.61 ± 2.76 ng/mL), indicating similar initial conditions for subsequent transport and clearance dynamics.

At 3 h post-dose, both groups exhibited a similar increase (UMFG: 151.40 ± 6.36 ng/mL; UTEV: 144.71 ± 8.13 ng/mL), suggesting that early-time systemic appearance is not strongly suppressed by the lipid shell (Fig. 3). In a transport interpretation, this implies that the shell does not impose an impenetrable barrier; rather, it modulates permeability while allowing a comparable early flux component.

A clear formulation-dependent separation emerged at later time points. At 6 h, UMFG relaxed toward baseline (109.01 ± 3.13 ng/mL), whereas UTEV maintained a higher concentration (123.84 ± 4.58 ng/mL). This persistence remained evident at 24 h (UMFG: 102.93 ± 3.04 ng/mL; UTEV: 114.87 ± 2.87 ng/mL). Collectively, the time-course indicates that the primary effect of the amphiphilic coating is not enhancement of the early peak but stabilization of the mid-to-late time tail. Within a core–shell framework, this behavior is consistent with reduced effective permeability $P = D_{eff}/\delta$ . and/or broadened release-time statistics arising from heterogeneous shell microstructure.

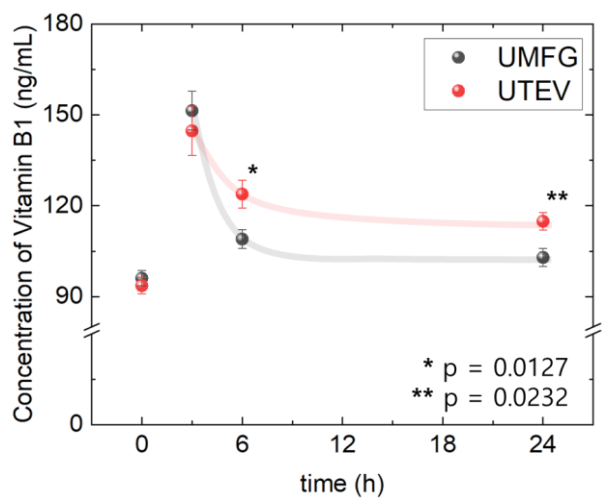

**Fig. 3.** Serum vitamin B1 time course (mean ± s.e.m.) following oral administration of UMFG (uncoated) and UTEV (GFAE-coated). Early-time appearance is comparable, while UTEV exhibits enhanced late-time persistence consistent with reduced permeability and broadened release-time statistics.

**Table 2.** Individual serum vitamin B1 concentrations (ng/mL) in mice following oral administration of UMFG or UTEV.

| Group | No. | Concentration (ng/ml) | | | |
|---|---|---|---|---|---|
| | | 0 h | 3 h | 6 h | 24 h |
| UMGF | 1 | 95.90 | 145.77 | 115.76 | 105.37 |
| | 2 | 99.13 | 158.61 | 101.92 | 111.47 |
| | 3 | 103.71 | 160.90 | 109.06 | 106.14 |
| | 4 | 81.02 | 116.10 | 90.19 | 84.03 |
| | 5 | 100.70 | 141.14 | 111.40 | 106.73 |
| | 6 | 97.26 | 147.84 | 115.37 | 104.10 |
| | 7 | 99.82 | 171.62 | 113.69 | 108.12 |
| | 8 | 91.14 | 169.20 | 114.67 | 97.46 |
| | Average | 96.09 | 151.40 | 109.01 | 102.93 |
| | S.E. | 2.52 | 6.36 | 3.13 | 3.04 |
| UTEV | 1 | 100.00 | 170.08 | 123.90 | 108.72 |
| | 2 | 93.78 | 141.37 | 121.59 | 107.92 |
| | 3 | 97.07 | 142.33 | 136.60 | 110.95 |
| | 4 | 85.64 | 108.26 | 105.77 | 124.04 |
| | 5 | 95.09 | 115.19 | 108.55 | 104.15 |
| | 6 | 90.32 | 166.58 | 119.69 | 123.64 |
| | 7 | 81.33 | 162.69 | 131.64 | 115.42 |
| | 8 | 105.68 | 151.20 | 143.01 | 124.07 |
| | Average | 93.61 | 144.71 | 123.84 | 114.86 |
| | S.E. | 2.76 | 8.13 | 4.58 | 2.87 |

In contrast, a clear separation between formulations emerged at later time points. At 6 h, serum vitamin B1 in the UMFG group declined toward baseline (109.01 ± 3.13 ng/mL), whereas UTEV maintained a higher concentration (123.84 ± 4.58 ng/mL). At 24 h, serum vitamin B1 remained near baseline in the UMFG group (102.93 ± 3.04 ng/mL), while UTEV continued to show a higher concentration (114.87 ± 2.87 ng/mL). Collectively, the time-course profile suggests that the primary formulation-dependent difference is not an increased early response but enhanced maintenance at mid-to-late time points, consistent with the intended sustained-release behavior of lipid coating. The detailed values are summarized in **Table 2**.

## 4.2 Baseline-normalized dynamics highlight stochastic broadening of release times

To isolate persistence relative to each animal's pre-dose level, serum concentrations were normalized to baseline (0 h) and expressed as a percentage of baseline (Fig. 4). Both UMFG and UTEV showed comparable baseline-normalized increases at 3 h, consistent with the similarity in absolute concentrations observed in the early regime. In contrast, UTEV exhibited significantly higher baseline-normalized levels at later times, with clear statistical separation at 6 h ($p$ = 0.00425) and 24 h ($p$ = 0.0116) (Fig. 4).

From a condensed-matter standpoint, the persistence enhancement is naturally interpreted as an ensemble effect: particle-to-particle variability in shell thickness $\delta$, connected hydrophilic fraction $\varepsilon$, and tortuosity $\tau$ yields a distribution of permeabilities and hence a broad distribution of characteristic release times. This broadening increases the probability mass of long-lived release events, producing a heavier late-time contribution to the measured signal under systemic clearance. In other words, the coating acts as a heterogeneous transport barrier whose disorder shifts release from a narrow time window into a wider distribution without substantially increasing the early-time flux.

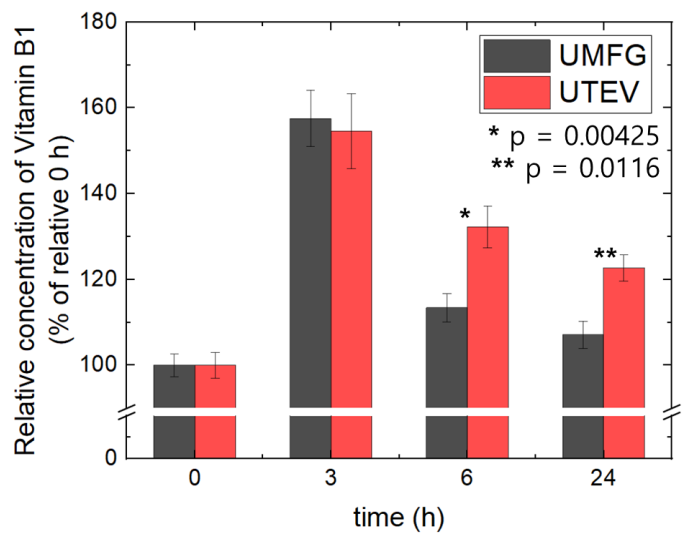

**Fig. 4**. Baseline-normalized serum vitamin B1 following UMFG or UTEV administration. Values are expressed as % of 0 h (mean ± S.E., n = 8). Significant separation at 6 h and 24 h indicates enhanced persistence and broadened release-time statistics for UTEV.

### 4.3 AUC enhancement arises from late-time tail contributions

Systemic exposure over the 0–24 h observation window was evaluated by calculating the area under the concentration–time curve, AUC(0–24 h). The AUC for UTEV was 2904.95 ± 71.61 ng/mL·h, compared with 2664.57 ± 71.96 ng/mL·h for UMFG (Fig. 5), corresponding to an approximately 1.09-fold higher AUC for the coated formulation. Because early-time concentrations were comparable between formulations, the increased AUC is attributed primarily to the late-time tail of the concentration profile, consistent with the sustained maintenance pattern observed at 6 h and 24 h. Within a coarse-grained interfacial-transport framework, this behavior implies a coating-induced shift of the effective release kernel $r_{rel}(t)$ toward longer times—arising from reduced effective permeability $P = D_{eff}/\delta$ , broadened permeability statistics due to shell heterogeneity, and/or stochastic trapping that generates longer-lived release events. Consequently, the amphiphilic lipid coating contributes to prolonged systemic availability of vitamin B1 under physiological clearance. Consistent with this interpretation, modified-release vitamin formulations have been reported to prolong exposure and increase AUC in human pharmacokinetic studies [16,17].

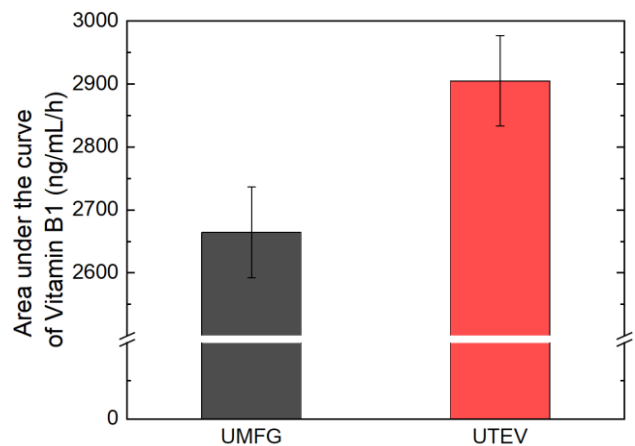

**Fig. 5.** AUC(0–24 h) of serum vitamin B1 concentration–time profiles after oral administration of UMFG or UTEV. The enhancement for UTEV is primarily attributed to the late-time tail rather than early-time peak amplification.

### 4.4 Limitations and Future Directions.

Several limitations constrain the present interpretation and point to clear next steps. First, vitamin B1 was quantified in serum by ELISA, which provides a robust comparative time-course but does not directly measure the biologically active intracellular pool (e.g., erythrocyte thiamine diphosphate, TDP). Future studies should incorporate HPLC-based TDP quantification in erythrocytes and tissue-level measurements to strengthen mechanistic and translational relevance. Second, the sampling schedule is sparse (0–24 h with four time points), limiting formal parameter extraction and obscuring early-time transport regimes; denser sampling at early times and an extended observation window would better constrain both absorption and long-time release tails. Third, the current study evaluates a single dose in one animal model; dose–response studies, repeated dosing, and food-effect testing are warranted to probe robustness under practical intake conditions. Finally, confirmatory human studies will be required to determine whether the observed persistence enhancement translates into improved functional outcomes and tolerability in real-world nutraceutical use.

From a condensed-matter/soft-matter perspective, a valuable direction is direct microstructure characterization of the amphiphilic shell (e.g., cryo-SEM, SAXS/DSC) combined with in vitro sink-release experiments. Such data would allow quantitative mapping from shell microstructure ($\varepsilon, \tau, \delta$) to D_eff, permeability distributions $P$, and potentially CTRW exponents ($\beta$), thereby connecting formulation design variables to macroscopic transport observables.

## 5. Conclusion

In this study, we developed a lipid-coated thiamine formulation (UTEV) using a glycerin fatty acid ester (GFAE) coating strategy and evaluated its in vivo time-course performance in male C57BL/6 mice. Reframing the formulation as a core–shell soft-matter system highlights the central role of interfacial diffusion control in governing release dynamics. Following a single oral administration, UTEV produced an early post-dose response comparable to the uncoated comparator (UMFG), indicating that the coating does not markedly amplify the early peak. In contrast, UTEV demonstrated improved late-time persistence, with baseline-normalized analysis showing significantly higher relative serum vitamin B1 levels at 6 h (p = 0.00425) and 24 h (p = 0.0116) compared with UMFG. Integrated exposure over the observation window was also higher with UTEV (AUC(0–24 h): 2904.95 ± 71.61 vs 2664.57 ± 71.96 ng/mL·h), consistent with a sustained maintenance pattern driven primarily by the late-time tail. Within a coarse-grained transport picture, these observations are consistent with a reduced effective interfacial diffusivity (and thus permeability) and broadened release-time statistics arising from heterogeneous amphiphilic shell microstructure, enabling stochastic release without suppressing early availability. Overall, these findings support GFAE-based amphiphilic lipid coating as a physically interpretable and practical formulation strategy to shift thiamine delivery from an absorption-centric profile toward a persistence-centric profile, with potential utility in nutraceutical and functional food applications.

Future work combining direct shell microstructure characterization (e.g., cryo-SEM, SAXS/DSC) with in vitro sink-release assays and distribution-based fitting of release-time statistics (e.g., lognormal versus CTRW-like heavy tails) will enable quantitative mapping from microstructure parameters ($\varepsilon, \tau, \delta$) to transport observables ($D_{eff}, P$) and potentially to CTRW exponents ($\beta$).

## DATA AVAILABILITY

Interested researchers may contact the Subjects Division representative for our study at unexa301@gmail.com to request access to the datasets. Access to data will be restricted to those who complete data sharing agreements.